\documentclass[fleqn,usenatbib]{mnras}

\usepackage[T1]{fontenc}

\DeclareRobustCommand{\VAN}[3]{#2}
\let\VANthebibliography\thebibliography
\def\thebibliography{\DeclareRobustCommand{\VAN}[3]{##3}\VANthebibliography}

\usepackage{graphicx}	
\usepackage{amsmath}	
\usepackage{amssymb}	
\usepackage[varg]{txfonts}
\usepackage{natbib}
\usepackage{epstopdf}
\usepackage{float}
\usepackage{etoolbox}

\usepackage{upgreek}

\usepackage{enumitem}
\setlist[itemize]{noitemsep, topsep=0pt}

\usepackage{accents}

\usepackage[greek,english]{babel}

\usepackage{dcolumn}
\newcolumntype{L}{D{.}{.}{0,5}}

\title[Skewness in the WR+BH Binary IC10 X-1]{Accreting Black Holes Skewing and Bending the Optical Emission from Massive Wolf-Rayet Companions---A Case Study of IC10 X-1}

\author[Bhattacharya et al.]{Sayantan Bhattacharya,$^{1,2}$\thanks{E-mail: sayantan34@gmail.com -- Corresponding author}
Dimitris M. Christodoulou,$^{2}$\thanks{E-mail: dmc111@yahoo.com}
André-Nicolas Chene,$^{3}$
Silas G. T. Laycock,$^{1,2}$
\newauthor{Breanna A. Binder,$^{4}$ and Demosthenes Kazanas$^{5}$}
\\
$^{1}$Department Of Physics \& Applied Physics, University Of Massachusetts, Lowell, MA, 01854, USA\\
$^{2}$Lowell Center For Space Science \& Technology, University Of Massachusetts, Lowell, MA, 01854, USA\\
$^{3}$Gemini Observatory/NSF’s NOIRLab, 670 N. A‘ohoku Place, Hilo, Hawai‘i, 96720, USA\\
$^{4}$Department of Physics \& Astronomy, California State Polytechnic University, Pomona, CA 91768, USA\\
$^{5}$NASA Goddard Space Flight Center, Astrophysics Science Division, Code 663, Greenbelt, MD 20771, USA
}

\date{Accepted XXX. Received YYY; in original form ZZZ}

\pubyear{2022}

\begin{document}
\label{firstpage}
\pagerange{\pageref{firstpage}--\pageref{lastpage}}
\maketitle

\begin{abstract}
We present a statistical analysis of the \ion{He}{ii} 4686 emission line in the spectra of the black hole and Wolf-Rayet (WR) star of the high-mass X-ray binary IC10 X-1. This line is visibly skewed, and the third moment (skewness) varies with the binary's orbital phase. We describe a new method of extracting such weak/faint features lying barely above a noisy continuum. Using the moments of these features, we have been able to decompose these skewed lines into two symmetric Gaussian profiles as a function of the orbital phase. The astrophysical implications of this decomposition are significant due to the complex nature of wind-accretion stream interactions in such binary systems. Previous studies have already shown a 0.25 phase lag in the radial velocity curve of the star and the X-ray eclipse, which indicates that the \ion{He}{ii} emitters might be in the stellar wind, hence not tracing the star's orbital motion. Results from this work further suggest the existence of two separate emitting regions, one in the stellar wind in the shadow of the WR star, and another in the accretion stream that impacts the black hole's outer accretion disk; and the observed skewed \ion{He}{ii} lines can be reproduced by superposition of the two corresponding time-dependent Gaussian emission profiles.

\end{abstract}




\section{Introduction}
Spectroscopic observations of Wolf-Rayet (WR) stars have always revealed interesting physics about their environments. WR stellar spectra are dominated by multiple emission lines (He, N, C, and O) due to a massive and optically thick wind emanating from the WR star \citep{,wolf1867spectroscopie,crowther2007review}. Analyzing these spectral features serves the versatile purpose of studying the star itself, its companion, and their interactions. In this work, we are studying a member of this rare subset of binaries, IC10 X-1, an eclipsing high-mass X-ray binary (HMXB) consisting of a black hole (BH) of estimated mass $\sim$15-35$M_\odot$ and a WN3-type WR star \citep{ prestwich2007orbital, laycock2015chandra, laycock2015revisiting}. \citet{clark2004wolf} determined [MAC92] 17A to be the donor star in the IC10 X-1 system and identified it as a WNE star situated in a crowded field with three other stars to within 0.3$^{\prime\prime}$-0.4$^{\prime\prime}$ of the X-ray source. In this study, a model spectrum \citep{hillier1998treatment} was used to find the stellar parameters: T$_*$ = 85,000 K, 
 log(L/\(\textup{L}_\odot\)) = 6.05, $\dot{M}$= 4 $\times$ 10$^{-6}$ $\textup{M}_\odot$\,yr$^{-1}$, and v$_\infty$ = 1750 km~s$^{-1}$. Several early-type WR stars in the Small Magellanic Cloud (SMC) (similar in metallicity to IC10) have similar properties to [MAC92] 17A. On the other hand, the association of [MAC92] 17A with a compact object makes it more similar to X-ray sources, such as Cyg X-3 and NGC300 X-1, rather than isolated or non-compact binary WR stars. \citet{silverman2008} monitored this system with 10 Keck spectra and determined a radial velocity of 370$\pm$20 km s$^{-1}$. This resulted in a BH mass in the range of 23.1 $\pm$ 2.1 $\textup{M}_\odot$ to 32.7 $\pm$ 2.6 $\textup{M}_\odot$, making IC10 X-1 the most massive stellar-mass BH at that time.

In HMXBs, the compact object (possibly a BH in our case) accretes material from the stellar wind of the massive companion (the WR star in IC10 X-1), and in turn, it emits X-ray photons. Most of the stellar wind is highly ionized, except for the shadow region behind the WR star with respect to the BH. This phenomenon is very clearly observed in the spectra of these systems. The \ion{He}{ii} 4686 line presumably originates from the relatively smaller and less ionized shielded region behind the WR star \citep{van1992infrared,van1993spectroscopic,van1996wolf}. The radial velocity curve determined from this configuration \citep{silverman2008} tracks the wind's motion rather than the stellar orbital motion \citep{laycock2015chandra}, hence we observe a 0.25 phase lag in the radial velocity (RV) curve, rendering the BH's mass determination inaccurate. In the absence of a well-determined compact object mass, the other parameters (orbital period of the system, eclipse duration, and the stellar parameters of the WR star) can be used to explore more plausible solutions \citep{binder2021wolf}. This mass conundrum also opens the door for consideration of a low-mass BH or even a neutron star (NS) companion, as shown in Table 1 of \citet{laycock2015revisiting}. Although the \ion{He}{ii} emission line is not useful in determining accurate Keplerian binary parameters, it can provide solid information about the stellar wind itself, viz., its ionization structure and velocity distribution. In this work, we have studied the time evolution of statistical moments (primarily up to the fourth moment) of the \ion{He}{ii} emission line and we have deciphered the physics governing the outflow from the orbiting WR+BH binary in IC10 X-1.

The WR wind has a high mass loss rate and outflowing velocity \citep{castor1975radiation, prinja1990terminal}, and the accretion by the compact object varies drastically around the stellar orbit \citep{tutukov2016binary}. The velocity vectors of multiple wind components imprint their signatures on the optical emission lines. The pronounced skewness of the \ion{He}{ii} 4686 emission line and its variation over the binary orbit can be used to understand the accretion (BH) - wind (WR star) interaction. The data can even be used to determine the locations of individual \ion{He}{ii} emitters around the binary orbit, as we show in this work.

Comparisons of the WN star in IC10 X-1 with WR stars, especially the WN population in the Large Magellanic Cloud \citep{hainich2014wolf} and the SMC \citep{hainich2015wolf}, can help us understand the effects of the compact object's presence in more detail. Such studies derive fundamental stellar parameters of WN stars including the distribution of terminal velocities ($v_\infty$). The value of $v_\infty$ of our WN star in IC10 X-1 is comparable to the results of the above-mentioned studies; 
a more complete comparison of other parameters using fits to radiative transfer models requires much higher-quality spectra. 

An emission line profile can be mathematically approximated by a Gaussian function. Depending on the movement of the line's centroid, we can determine the velocity of the emitting regions using Doppler shifts. This is a very well-known phenomenon, but when the observed emission line is skewed, the measured velocity could be a representation of multiple superposed velocity vectors. When we decompose such a skewed Gaussian line profile into two (or more) unskewed components, then the velocities of individual components can be determined. Such asymmetry in the profile could be caused by an inhomogeneous stellar wind from the WR star, as shown by \cite{hamann1998spectrum}; in their work, the inhomogeneity is caused by clump formation inside the wind. Profile asymmetry may also be caused by the presence of a compact object around its orbit, and we have tried to model this scenario in the present work; we have formulated a new analytic method to decompose the skewed emission lines into distinct Gaussian components, and we discuss the physics of two such components as they evolve with orbital phase.

The outline of the paper is as follows: In \S~\ref{procedure}, we describe the archival data used in this work and the method used to detect and analyze the especially weak \ion{He}{ii} 4686 emission lines at different phase bins. In \S~\ref{decomposition}, we describe the analytic results of decomposing a skewed Gaussian profile into two unskewed Gaussian components. In \S~\ref{discuss}, we discuss the astrophysical implications of this decomposition for IC10 X-1, and we conclude in \S~\ref{sum} with a summary of our results.    

\section{Observations and Reduction Methods} \label{procedure}

IC10 X-1 has been observed by the {\it GEMINI-North/GMOS} telescope through multiple observation campaigns over a long time span (2001-2019), and we have analyzed all of these archival data. The spectral data are summarized in Table \ref{tab:gmos_data}, all collected in the {\it MOS} mode and archived by \cite{spec_data} \citep[see also][]{bhattacharya2023probing}. All the spectra were obtained using the B600 grating in the {\it MOS} mode, most of the time with a slit-width of $1.0^{\prime\prime}$. The B600 grating has a resolving power R=1688 at the blaze wavelength of 461 nm. The 0.5 \AA/pixel dispersion results in a $\approx$ 32 km/s velocity resolution per pixel at 4686~\AA. Spectra from different observation campaigns with different observing conditions have been phasewise stacked, hence the quality varies significantly in each phase bin. In particular, spectra around phase $\phi = 0.7$ are of very low quality, and the \ion{He}{ii} line is barely detected above the noise.

The \textit{gemini} package in \textit{iraf} was used to process the raw data. Steps described in \textit{gmosexample} have been followed to perform standard calibrations and extraction of the one-dimensional spectra. No preliminary sky subtractions were performed at this stage because of the faintness of the source. The background was subtracted by a new procedure described in detail in \S~\ref{process} below. Conventional techniques do not "see" the \ion{He}{ii} 4686 emission line protruding above the noise. The new procedure starts by taking into account the asymptotic decay of the line into the surrounding background.

In the process, we obtained a significant number of spectra (52); yet, the \ion{He}{ii} line was not detected in most of them individually; hence, we had to stack them to amplify the signal. The spectra were stacked phase-wise to increase the signal-to-noise ratio in each phase bin. Two binning schemes were used, one involving 10 bins of width $\Delta\phi = 0.1$, and another with 4 bins of width $\Delta\phi=0.25$. Then, the two data sets were analyzed together in order to look for variations in the moments of the \ion{He}{ii} 4686 line as a function of the orbital phase. 

\begin{table}
    \centering
    \caption{\textit{GEMINI/GMOS} Available Archival Data}
    \begin{tabular}{l c c c c}
        \hline 
        Program  & Slit & Start & End   & Number  \\ 
            ID   & Width ($^{\prime\prime}$) & Date  & Date  &  of Spectra \\
        \hline
        \noalign{\vskip 0.5mm}
        GN-2001B-Q-23 & 0.8 & 2001-12-22 & 2002-01-17 & 5 \\ 
        GN-2004B-Q-12 & 1.2 & 2004-07-16 & 2004-08-13 & 9 \\ 
        GN-2010B-Q-58 & 0.75 & 2010-09-02 & 2010-09-07 & 6 \\ 
        GN-2017B-Q-20 & 1.0 & 2017-11-10 & 2017-11-15 & 7 \\ 
        GN-2018B-Q-127 & 1.0 & 2018-12-13 & 2019-01-04 & 10 \\ 
        GN-2018B-Q-127& 1.0 & 2019-07-02 &2019-07-04 & 15 \\
         \hline
    \end{tabular}
    
    \label{tab:gmos_data}
\end{table}
\subsection{New Data Fitting Procedure for a Noisy Distribution with a Weak Embedded Signal}\label{process} 

At faint stellar magnitudes, the {\it Gemini/GMOS} spectra are dominated by strong emission lines due to the intervening atmosphere and by the noise that makes it hard to find out where the continuum lies around the location of a weak optical emission line such as the \ion{He}{ii} 4686 line coming from the WR star or its wind in IC10 X-1 and similar WR binary systems. The usual procedures for spectrum decontamination do not work, even after piling up many spectra within each phase bin to amplify the signal. So, we had to devise an alternative procedure capable of fitting the \ion{He}{ii} line along with its true continuum. (An example, the $\phi=0$ case with a bin width of $\Delta\phi = 0.25$, is illustrated in Figure~\ref{fig:comb_figs}.) The method proceeds in the following steps:

\begin{figure*}
    \centering
    \includegraphics[trim=0cm 5cm 0cm 5cm, clip, width=\textwidth]{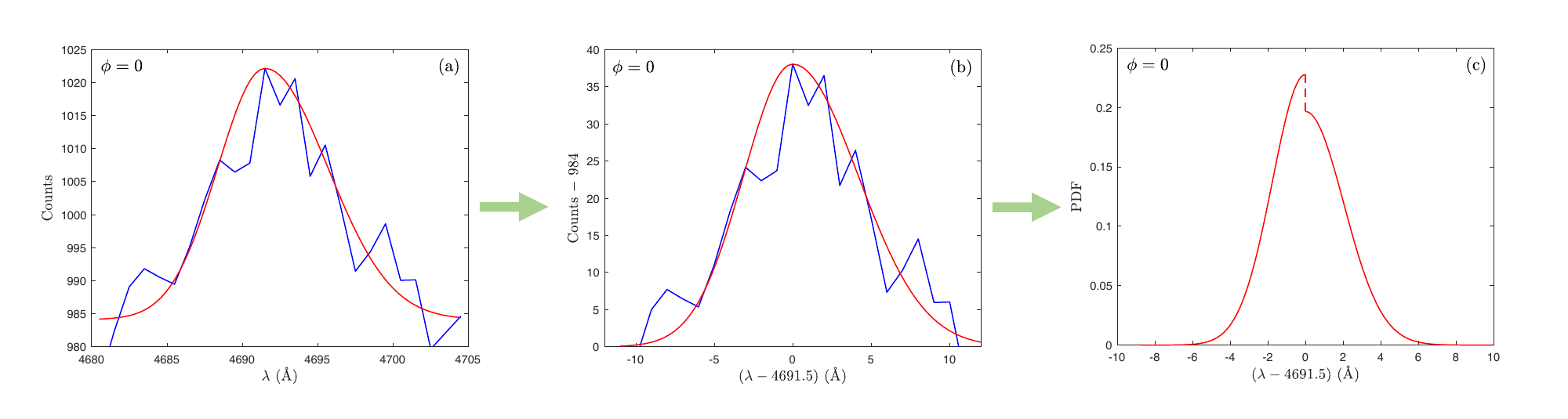}
    \caption{Models (in red) fitted to the \ion{He}{ii} 4686 data (in blue) in a sequence of steps at phase $\phi = 0\pm0.125$. (a) Exponential fits on either side of $\lambda = 4691.5$ \AA~ using the same maximum (1022 counts). True continuum (984 counts) is determined by the fit and the nearby background on the left side of the line. (b) Exponential fits on either side of the maximum (38 counts) after a vertical shift of $-984$ counts and a horizontal shift of $-4691.5$ \AA. (c) Half-Gaussian fits on either side of the mean ($\mu = 0$). Integrating this synthetic model, we obtain the higher-order moments of the combined distribution (top row with $\phi = 0.0$ in Table~\ref{t3} below).}
    \label{fig:comb_figs}
\end{figure*}

(a)~Fitting starts with a visual inspection of the combined spectrum in each phase bin. We determine through experimentation the footpoints of the line and the level $C_1$ of the apparent continuum on the left side of the line. The left side is used because there are no important lines except N III 4630/43, which is in this case buried in the jitter, and we avoid that region in all cases. The right side is inappropriate for use because the apparent continuum is very much raised due to the presence of strong nearby emission lines such as the H$\beta$ 4861 line. The jitter of a long segment on the left side is not fitted by a polynomial function, as is usually done. That would depress the apparent continuum significantly. Instead, we find the maximum number of counts in the jitter near the left side of the left footpoint of the \ion{He}{ii} 4686 line, and we adopt this value for $C_1$. Note that this value also includes the asymptotic left tail of the \ion{He}{ii} line.

(b)~We fit the left side of the \ion{He}{ii} line with a decaying exponential function, $C(\lambda) = L_1 + (C_{\rm max}-L_1)\exp(-|\lambda - \mu|/(2\varv))$, where $(\mu, C_{\rm max})$ is the mode and $\varv$ is a free parameter resembling variance; and we determine $L_1$, the asymptotic value of the line that contributes to the region of the apparent continuum. Then, the true continuum lies at a level of $C_0 = C_1 - L_1$. In this region devoid of other lines, if $C_1 > L_1$, we obtain a reasonable value of true continuum $C_0$. In the few cases in which $L_1\gtrsim C_1$ (differing by no more than 3-5 counts), we reset $C_0 = L_1$ (because this is where the tail of the line is headed after all; and then, it specifies the true continuum all by itself). Such cases arise from the poor quality of the spectra that we are analyzing, but the differences in counts are too small to be of particular significance. On the other hand, any traditional attempt to use the mean jitter level (ignoring background flaring and the tail of the line) for removing the ``visible continuum'' is doomed to failure---the weak \ion{He}{ii} line signal is washed out by the noise.

(c)~Having determined the true continuum $C_0$, we proceed to fit both sides of the \ion{He}{ii} line, again with exponential functions that have the appropriate asymptotic behavior $C_0$ (Figure~\ref{fig:comb_figs}a), and we determine their pseudo-variances, say $\varv_{_{\rm LS}}$ and $\varv_{_{\rm RS}}$ (on the left side and the right side, respectively). In the few cases with $C_0 = L_1$, then $\varv_{_{\rm LS}} = \varv$.

(d)~The newly fitted curves are still not Gaussian because their amplitudes and areas are not consistent (Figure~\ref{fig:comb_figs}b). Thus, we reset the amplitudes so that the area under each half-curve is 1/2; and we fit again the two half-profiles, this time with actual half-Gaussian functions, to determine their true variances, say $V_{\ell}$ and $V_{r}$. For convenience, we also shift $F(\lambda)$ in wavelength space, so that its mode is located at $\mu = 0$.

(e)~We concatenate the two half-Gaussians into one distribution function $F(\lambda)$ (Figure~\ref{fig:comb_figs}c). It is easy to verify by numerical integration that the total area under the curve is equal to 1, and we did so.

(f)~Now, we are ready to extract the \ion{He}{ii} signal out of the noisy spectral data. The best-fitted profile is not noisy at all, although we can assign typical error bars to its values by standard uncertainty propagation analysis. We find that the relative errors do not exceed $\pm10\%$ in all observed points. We choose a grid spacing of 1~m\AA (so that numerical errors are negligible), and we generate synthetic data points from the distribution $F(\lambda)$ out to $-5\sqrt{V_{\ell}}$ on the left and $+5\sqrt{V_{r}}$ on the right of the mode $\mu = 0$. The large number of data points generated guarantees that all corrections for bias will be quite small. Nevertheless, we apply the small bias corrections \citep{joa98} to the moments discussed below. 

\subsection{The Distribution Function $F(\lambda)$}

The final distribution function $F(\lambda)$ is obviously asymmetric with a discontinuity at $\mu = 0$ (Figure~\ref{fig:comb_figs}c). This does not prevent us from determining its moments by direct integrations. We computed and inspected the results up to the $8^{\rm th}$ normalized moment $u(8)/[u(2)]^4$, and we confirmed that we see a real signal in all of these moments (for example, moment $u(8)/[u(2)]^4$ must be significantly larger than $7!! = 105$ --- indeed, it is, in all phase intervals under consideration).

We also investigated whether skewness could arise from instrumental effects. We used a WN star model spectrum \citep[PoWR model;][]{grafener2002line}, and we convolved it using the instrumental line spread function. We did not find any significant skewness in this synthetic model of the \ion{He}{ii} 4686 line. Hence, we can be confident that the observed skewness is due to physical causes associated with the IC10 X-1 compact binary system.

In this work, we are primarily interested in the lower three normalized moments of the distribution function $F(\lambda)$ besides the mean; specifically, the $2^{\rm nd}$ (variance ${\cal V}$), $3^{\rm rd}$ (skewness ${\cal S}$), and $4^{\rm th}$ (kyrtosis\footnote{This ought to be the Latinized spelling of the Greek word ``\textgreek{k\'urtosis}'' that translates to ``curvature'' or ``bending," also used in the title; we note that Greek authors consistently use ``kyrtosis" (with a "y") in the literature.} 
 ${\cal K}$) moments. We analyze these moments analytically in \S~\ref{decomposition}, where we decompose the signal into two partially overlapping Gaussian components, and we follow their evolution as a function of orbital phase.

\section{Skew-Kyrtic Model Decomposition into two Pure Gaussian Components}\label{decomposition}

We assume that the best-fit model for the \ion{He}{ii} line of IC10 X-1 is a unimodal superposition of two Gaussians, each of which carries a weight of $p=1/2$, and that are located on either side of the observed mode $\mu$. In this pilot study, we have no reason to favor one distribution against the other by using different weights, in which case we would have to include the 5$^{\rm th}$ moment too, and then solve a 9$^{\rm th}$-order polynomial equation \citep{pea94,eve81}.

\subsection{Toward an Analytic Solution}\label{toward}

With the parameters $\mu, {\cal V}, {\cal S}$, and ${\cal K}$ determined by the above procedure in each phase bin around the orbit of the binary system, we proceed to decompose the skewed and kyrtic distribution function $F(\lambda)$ (see, e.g., Figure~\ref{fig:comb_figs}c) into two partially overlapping Gaussians with means $\mu_i$, variances $V_i$, and corresponding normalized moments $S_i=0$ and $K_i=3$, where $i=1, 2$, respectively, and $\mu_1 \geq \mu_2$ by design. Having adopted $p=1/2$ for the weights of the mixture, we are called to solve only a cubic polynomial equation, an analytically tractable endeavor, which we describe below.

Considering the first four equations in the nonlinear system of equations~(2.10) given by \cite{eve81} and using $p=1/2$, $S_i=0$, and $K_i=3$, we reduce the Gaussian solution set \{$\mu_\pm, V_\pm$\} (where $\pm$ corresponds to $i=1, 2$, respectively) to obeying the following system of equations:
\begin{equation}
    \begin{array}{cll}
0         &=& \delta_1 + \delta_2 \\
2{\cal V} &=& V_1 + \delta_1^{~2} + V_2 + \delta_2^{~2} \\ 
2{\cal S} &=& \delta_1\left(3V_1+\delta_1^{~2}\right)
             + \delta_2\left(3V_2+\delta_2^{~2}\right) \\ 
2{\cal K} &=& 3V_1^{\,2} + 6V_1\delta_1^{~2} + \delta_1^{~4}
         + 3V_2^{\,2} + 6V_2\delta_2^{~2} + \delta_2^{~4} \\
    \end{array}
    ,
\end{equation}
where
\begin{equation}
\delta_i = \mu_i - \mu~~~{\rm for}~~i=1, 2\, .
\end{equation}

\textbf{The above equations can be reduced to a system involving a cubic equation for a new variable $x\geq 0$ (defined in equation~(\ref{x1}) below), viz.}
\begin{equation}
6x^3 + 3\left({\cal K} - 3\right) x - {\cal S}^2 = 0
\, ,
\label{x3}
\end{equation}
\begin{equation}
\mu_\pm = \mu \,\pm\, c
\, ,
\label{mu12}
\end{equation}
\begin{equation}
c = \sqrt{{\cal V}x\,} = {\cal D}\sqrt{x\,}
\, ,
\label{c12}
\end{equation}
and
\begin{equation}
V_\pm = {\cal V}\left(1 - x \,\pm\, \frac{\cal S}{3\sqrt{x}}\right)
\, ,
\label{sig12}
\end{equation}
where ${\cal D}\equiv {\cal V}^{1/2}$ is the standard deviation of the original input distribution. These equations also imply some important auxiliary relations between parameters, viz.
\begin{equation}
\mu_1 + \mu_2 = 2\mu\, ,~~~~\mu_1 - \mu_2 = 2c \ \geq \ 0 \, ,
\label{mus}
\end{equation}
and 
\begin{equation}
x = (c/{\cal D})^2 \ \geq \ 0 \, .
\label{x1}
\end{equation} 
We see now that $x$ is dimensionless, since $c$ and ${\cal D}$ have dimensions of [length].

\begin{figure}
\begin{center}
    \leavevmode
      \includegraphics[trim=0 0cm 0 0cm, clip, angle=0,width=9 cm]{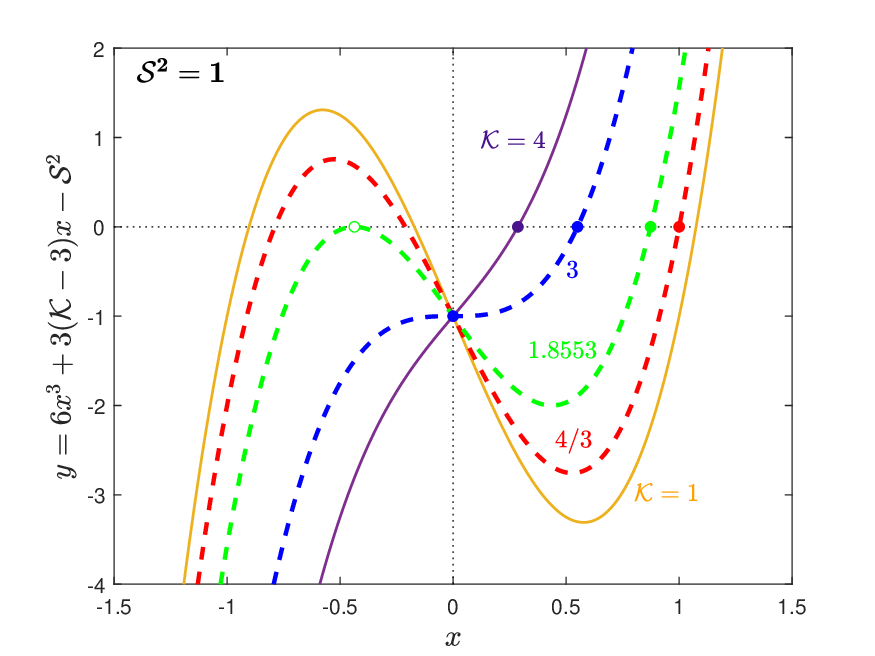}
\caption{The cubic function $y(x)$ is plotted versus $x$ for ${\cal S}^2=1$ and various values of the kyrtosis ${\cal K}$. Solid dots mark the zeroes in the interval of interest $x\in(0, 1]$. For ${\cal K}< 4/3$, the positive zeroes move out beyond $x=1$. For ${\cal K} = 1.8553$, the equation $y(x)=0$ has a double root at $x=-0.43679$. For ${\cal K} = 3$, the function has one real zero at $x=0.55032$, and an inflection point at point $(0, -1)$. For ${\cal K} > 3$, the function has one real zero in the interval $x\in(0, 0.55032)$. The curves for ${\cal K}\leq 2$ (and zeroes $x \geq 0.83624$) have no physical significance for ${\cal S}^2 = 1$ (equation~(\ref{pearson_sk}) below).
\label{fig_fofx}}
  \end{center}
\end{figure}

Equation~(\ref{x3}) can be solved analytically for the nontrivial case, where ${\cal S}\neq 0$. Depending on the parameters ${\cal S}$ and ${\cal K}$, it has either one real positive root or three real roots, two of which are negative. Thus, a solution to our problem (a positive real root) always exists and it is unique. Figure~\ref{fig_fofx} shows an example in which we set ${\cal S}^2 =1$ in equation~(\ref{x3}) and we plotted the cubic polynomial $y(x) = 6x^3 + 3\left({\cal K} - 3\right) x - 1$ for various values of ${\cal K}$. The caption describes the various cases involving the zeroes of $y(x)$. 

The general equations that we used to obtain the critical points shown in Figure~\ref{fig_fofx} are obtained from the cubic function of equation~(\ref{x3}), viz.
\begin{equation}
y(x) = 6x^3 + 3\left({\cal K} - 3\right) x - {\cal S}^2\, ,
\label{yofx}
\end{equation}
as follows: 
\begin{itemize}
\item[(a)]The critical point (1, 0) is obtained directly from $y(1)=0$. Then, we find that it occurs for ${\cal K}=1 + {\cal S}^2/3$ (${\cal K}=4/3$ in Figure~\ref{fig_fofx}). 

\item[(b)]The inflection point at $x=0$ is obtained from $y^{\prime\prime}(0) = 0$. We also find that $y^\prime(0) = 0$ for ${\cal K}=3$, as seen in Figure~\ref{fig_fofx}. 

\item[(c)]The real double root at negative $x$-values is obtained by solving simultaneously the system of equations $\left\{\,y(x)=0,\,y^{\prime}(x)=0\,\right\}$. We then find that $x = -({\cal S}^2/12)^{1/3}$ and ${\cal K}=3 - (3{\cal S}^4/2)^{1/3}$ ($x=-0.43679$ and ${\cal K}=1.8553$ in Figure~\ref{fig_fofx}).

\item[(d)]The zero of the ${\cal K}=3$ case is obtained easily from $6x^3 - {\cal S}^2=0$. We find that $x = ({\cal S}^2/6)^{1/3}$ ($x=0.55032$ in Figure~\ref{fig_fofx}).

\item[(e)]The zero of the ${\cal K}=2, \,{\cal S}^2 = 1$ case ($x=0.83624$), given in the caption of Figure~\ref{fig_fofx}, is found analytically by solving the particular cubic equation $6x^3 -3x -1 = 0$.
\end{itemize}

\subsection{The Analytic Solution}

The ``discriminant'' of equation~(\ref{x3}) takes the form
\begin{equation}
d = 6\left({\cal K}-3\right)^3 + 9{\cal S}^4
\, ,
\label{bigd}
\end{equation} 
and $d>0$ for {\it leptokyrtic} distributions (``slender'' ones with ${\cal K}>3$) and {\it mesokyrtic} distributions (``middle'' ones with ${\cal K}=3$), such as the those we obtained for the \ion{He}{ii} line of IC10 X-1. For $d<0$, the above system of equations has two negative real roots and one positive real root. The critical case $d=0$ corresponds to the double-root case described in \S~\ref{toward}, item~(c) above. In such {\it platykyrtic} (``broad'' top, light tails) cases, the signal cannot be composed of a mixture of two Gaussians---no matter how the two Gaussians are arranged, close or far apart, they cannot get both the top of the mixture to be broad and the tails to be thin.

When $d > 0$ in equation~(\ref{bigd}) (i.e., for ${\cal K} > 3 - (3{\cal S}^4/2)^{1/3}$; cf. \S~\ref{toward}, item~(c) above), another intermediate variable appears to dominate the solution, viz.
\begin{equation}
{\cal Q} = 3{\cal S}^2 + \sqrt{d\,}
\, ,
\label{qsol}
\end{equation} 
so that the solution of equation~(\ref{x3}) takes the relatively compact form
\begin{equation}
x = \left(\frac{\cal Q}{36}\right)^{1/3} - \,~\left(\frac{({\cal K}-3)^3}{6{\mskip0.5\thinmuskip}{\cal Q}}\right)^{1/3}
\, ,
\label{sol}
\end{equation}
where, in general, we expect that $0 < x \leq ({\cal S}^2/6)^{1/3}$ (for ${\cal K}\geq 3$) in this work. Yet another, equivalent form of this solution turns out to be more convenient for computations:
\begin{equation}
x = \left( \frac{{\cal S}^2}{12} \right)^{1/3}\left[ \left(1 + \sqrt{1 + g\,}{\mskip0.75\thinmuskip}\right)^{1/3} - \left(\frac{g}{1 + \sqrt{1 + g\,}}\right)^{1/3} \right]
\, ,
\label{sol2}
\end{equation}
where
\begin{equation}
g = \frac{2}{3}\frac{({\cal K}-3)^3}{{\cal S}^4}
\, .
\label{sol22}
\end{equation} 
As stated above, here ${\cal S}\neq 0$, but the normalized kyrtosis could potentially take the value ${\cal K}=3$, in which case, we deduce that $g = 0$ and $x = ({\cal S}^2/6)^{1/3}$. We also find that, as $g\to 0$, the combination of terms in the square brackets tends to $2^{1/3}\left[1 - (g/4)^{1/3} + {\cal O}(g)\right]$ and then 
$$x\approx ({\cal S}^2/6)^{1/3}\,[1 - (g/4)^{1/3}]~.$$ 
Since $x<1$ in our problem, the normalized skewness should then be restricted approximately in the interval of ${\cal S}^2 < 6$ for ${\cal K}\approx 3$, but this range turns out to be too wide. A much more accurate constraint on ${\cal S}^2$ can be obtained from the momentous condition that 
\begin{equation}
{\cal S}^2 \,<\, {\cal K} - 1\, ,
\label{pearson_sk}
\end{equation}
proven by \cite{pea16,pea29} for any distribution. For ${\cal K} = 3$, this inequality predicts emphatically that ${\cal S}^2 < 2$, thus we expect that ${\cal S}\in(-1.414, +1.414)$ in our problem as well.

\subsection{Significance of the Output Parameters}\label{significance}

With $x$ determined from equation~(\ref{sol}) or equation~(\ref{sol2}), we can proceed to find the means $\mu_\pm$ (equation~(\ref{mu12})) and the variances $V_\pm$ (equation~(\ref{sig12})) of the two Gaussian signals in the mixture that produces the observed weak \ion{He}{ii} line in IC10 X-1. The main results are listed in Table~\ref{t1}, where the means of the two Gaussian lines are shown relative to the rest-frame wavelength of $4686~\AA$. Before we interpret these results, we should revisit the fundamental parameters discussed in the subsections above, and assign physical meaning to what is measured and derived in this analysis.

\begin{table}
\caption{\ion{He}{ii} Line Decomposition into Two Gaussian Signals}
\label{t1}

\addtolength{\tabcolsep}{-1pt}  
\hspace*{-\leftmargin}\begin{tabular}{rrrrrrr}
\hline
\noalign{\vskip 1mm}
$\phi$~~ & $\mu_1^{~\star}$ & $\mu_2^{~\star}$ & $V_1$~~ & $V_2$~~~ & $D_1$~~ & $D_2$~~ \\
\noalign{\vskip 0.5mm}
\hline
\noalign{\vskip 0.5mm}
0.1 & $8.12$ & $3.94$ & 5.24 & 15.08 &  2.29 & 3.88 \\
0.2 & $5.11$ & $-0.21$ & 5.39 & 22.30 &  2.32 & 4.72 \\
0.3 & $6.35$ & $1.66$ & 5.45 & 18.19 &  2.33 & 4.26 \\
0.4 & $-1.16$ & $-3.43$ & 25.89 & 23.74 &  5.09 & 4.87 \\
0.6 & $-5.28$ & $-7.72$ & 9.90 & 12.57 &  3.15 & 3.54 \\
0.7 & $0.77$ & $-0.32$ & 4.12 & 3.62 &  2.03 & 1.90 \\
0.8 & $3.76$ & $0.19$ & 3.42 & 10.72 &  1.85 & 3.27 \\
0.9 & $6.92$ & $5.47$ & 3.76 & 4.70 &  1.94 & 2.17 \\
1.0 & $7.09$ & $5.01$ & 6.35 & 4.30 &  2.52 & 2.07 \\ 

\hline

\end{tabular}
\vskip0.15cm
~~~~~$^{\star}$(Line Center) $- ~(4686~\AA)$.

\addtolength{\tabcolsep}{1pt}  

\end{table}

Pivotal parameter $c$ (equations~(\ref{mu12}) and~(\ref{c12})-(\ref{x1})) comes first in this deeper examination. According to equation~(\ref{mus}), $c = (\mu_1 - \mu_2)/2 \equiv \Delta\mu/2\geq 0$, thus $c$ is one-half of the separation $\Delta\mu$ of the two Gaussian means. Thus, we think of $c$ as unnormalized ``standard deviation'' of the two means, and we define a dimensionless {\it separation distance $D_{\mu}$ of the means} by
\begin{equation}
D_{\mu}\equiv \Delta\mu/(2{\cal D}) = c/{\cal D} = \sqrt{x\,}\, ,
\label{dmu}
\end{equation}
where ${\cal D} = {\cal V}^{1/2}$. Obviously then, the positive root of the cubic equation~(\ref{x3}) provides the separation distance squared ($x=D_{\mu}^{~2}$). Furthermore, since $x<1$, then $0\leq \Delta\mu < 2{\cal D}$ (although these limits are too wide), where we emphasize again that ${\cal D}$ is the standard deviation of the {\it original} mixed sample. Finally, we note that the inverse of $D_\mu$, i.e., $1/\!\sqrt{x}$, resembles the well-known ``coefficient of variation $CV$,'' which is equal to ${\cal D}/\mu$ for the original sample. 

A surprising realization concerning the equations of \S~\ref{toward} is that, just like equation~(\ref{dmu}), they all practically ``beg'' to be normalized to the variance ${\cal V}$ of the {\it original skewed and kyrtic} data set, and not to a combination of the derived Gaussian variances $V_1$ and $V_2$. This approach has not been implemented previously in a statistical context, probably because people do not generally feel comfortable mixing input and output parameters in the same normalized expression. We undertook a literature search, and we found one work \citep{cha10} in which the original sample's ${\cal V}$ was used as a normalization factor in the proposed bimodality index $k$, which in our notation would read\, $k = (V_1+V_2)/(2{\cal V})$.

\begin{figure}
\begin{center}
    \leavevmode
      \includegraphics[trim=0 0cm 0 0cm, clip, angle=0,width=9 cm]{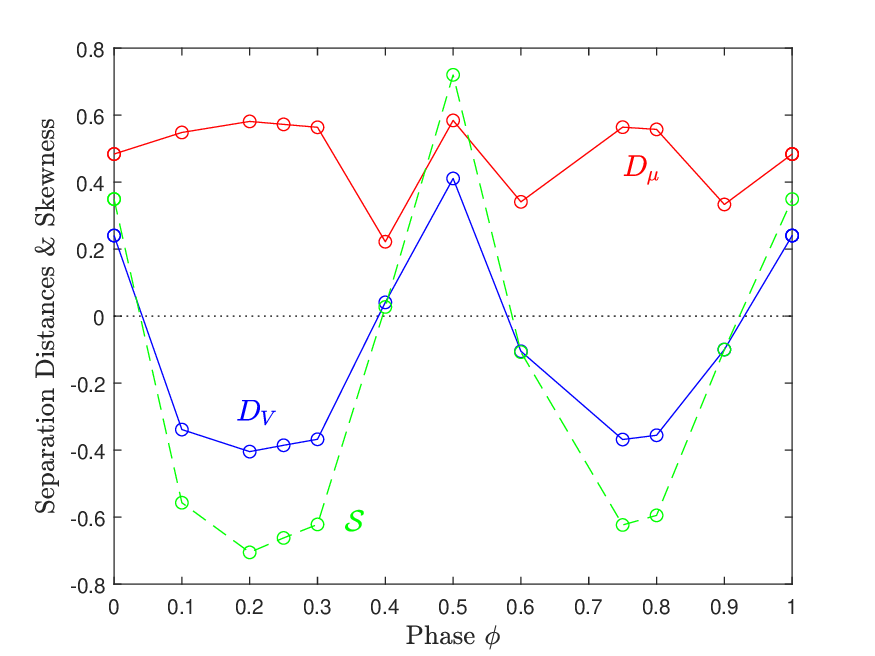}
\caption{Variation of separation distances $D_\mu$ and $D_V$ during one phase. Distance $D_V$ tracks closely the skewness ${\cal S}$ of the original distribution, which is also shown by dashed line segments. Distance $D_\mu$ is anticorrelated to ${\cal S}$, except in the blind sector around $\phi = 0.5$.
\label{fig_sep}}
  \end{center}
\end{figure}

Based on the above elaboration, we also define a dimensionless {\it separation distance $D_{V}$ of the variances} of the two Gaussians by
\begin{equation}
D_{V}\equiv \frac{V_1 - V_2}{2{\cal V}} \,=\, \frac{{\cal S}}{3\sqrt{x\,}} \,=\, \frac{{\cal S}}{3{\mskip0.3\thinmuskip}D_\mu}\, ,
\label{dv}
\end{equation}
and a dimensionless {\it bimodality index $B_{k}$ of the variances} \citep[akin to index $k$ of][]{cha10} of the two Gaussians by
\begin{equation}
B_{k}\equiv \frac{V_1 + V_2}{2{\cal V}} \,=\, 1 - x \,=\, 1 - D_{\mu}^{~2}\, .
\label{bk}
\end{equation}
The separation distances $D_\mu$ and $D_V$ (equations~(\ref{dmu}) and~(\ref{dv})) are plotted versus $\phi\in[0, 1]$ in Figure~\ref{fig_sep}, along with the skewness ${\cal S}$ of the original data set.

We note the following properties of the above relations:
\begin{itemize}
\item[(a)]~When $x=0$ (or $D_{\mu}=0$), then ${\cal S}\equiv 0$ necessarily, and the last two fractions in equation~(\ref{dv}) become indeterminate; then, $D_V$, which may be nonzero, can only be determined from the first equality in equation~(\ref{dv}). In this case, ~$\mu_1 = \mu_2 = \mu$, ~$\,\Delta\mu = c = 0$, ~and ~$B_{k} = 1$.

\item[(b)]~Equation~(\ref{dv}) shows that $D_V < 0$ when $V_1 < V_2$ (i.e., when the larger mean is associated with the smaller variance). Furthermore, ${\cal S}\propto (V_1 - V_2)/{\cal V}$, implying that the sign of $(V_1 - V_2)$ is determined from the sign of the skewness ${\cal S}$ of the initial distribution.

\item[(c)]~There is no need to define a bimodality index for the means $\mu_1$ and $\mu_2$, analogous to $B_k$ in equation~(\ref{bk}). Such an index would be equal to $\mu/{\cal D}$, the inverse of the coefficient of variation $CV\equiv {\cal D}/\mu$ that describes the original distribution. Naturally then, $CV$ also describes the same sample after its decomposition to two overlapping Gaussian distributions.

\item[(d)]~The inverse of the original $CV$ is the harmonic mean of the inverses of $CV_\pm = {\cal D}/\mu_\pm$, where $\pm$ corresponds to $i=1, 2$, respectively. That is, $1/CV_{+} + 1/CV_{-} = 2/CV$ for the inverse coefficients of variation of the two decomposed Gaussian distributions \textbf{with values given by} 
\begin{equation}
\frac{1}{CV_\pm} = \frac{1}{CV} \,\pm\, D_\mu \, .
\label{cvpm_dmu}
\end{equation}

\item[(e)]~Suppose one can determine $\mu_1$ and $\mu_2$ through some other means. For example, if the original distribution is clearly bimodal, then it seems reasonable to assume that the two visible modes represent the two Gaussian means. Then, there is no need to go through the long procedure that we described above. The separation distance $D_\mu$ is determined from the first equality in equation~(\ref{dmu}), and the variances are readily determined from equation~(\ref{sig12}) with $x=D_\mu^{~2}$, viz. 
\begin{equation}
V_\pm = {\cal V}\left(1 - D_\mu^{~2} \,\pm\, \frac{\cal S}{3{\mskip0.3\thinmuskip}D_\mu}\right)
\, .
\label{sig12again}
\end{equation}

\end{itemize}

\begin{table}
\caption{Diagnostic Tests for Mixture Modality}
\label{t2}

\hspace*{-\leftmargin}\begin{tabular}{cccccccc}
\hline
$\phi$ & $D_\mu$ & ~$D_V$ & $d_1$ & $d_2$ & $d_3$ & $d_4$ & $d_5$ \\
\noalign{\vskip 0.25mm}

\hline

\noalign{\vskip 0.75mm}

0.10&  0.548&   $-0.339$&  0.70&  1.31&  0.34&  0.70&   0.41 \\

0.20&  0.581&   $-0.404$&  0.80&  1.43&  0.38&  0.66&   0.46 \\

0.25&  0.572&   $-0.386$&  0.77&  1.40&  0.37&  0.67&   0.45 \\

0.30&  0.564&   $-0.368$&  0.74&  1.37&  0.36&  0.68&   0.43 \\

0.40&  0.222&   $+0.041$&  0.23&  0.46&  0.11&  0.95&   0.33 \\

0.50&  0.584&   $+0.411$&  0.81&  1.44&  0.38&  0.66&   0.46 \\

0.60&  0.341&   $-0.105$&  0.36&  0.73&  0.18&  0.88&   0.34 \\

0.75&  0.564&   $-0.368$&  0.74&  1.37&  0.36&  0.68&   0.43 \\

0.80&  0.558&   $-0.356$&  0.73&  1.34&  0.35&  0.69&   0.42 \\

0.90&  0.333&   $-0.100$&  0.35&  0.71&  0.18&  0.89&   0.34 \\

1.00&  0.484&   $+0.241$&  0.57&  1.11&  0.28&  0.77&   0.37 \\

\hline
\noalign{\vskip 0.5mm}
\multicolumn{3}{l}{Unimodality Condition} & $d_1\!\leq\!1$ & $d_2\!<\!2$ & $d_3\!\not\lesssim\!1$ & $d_4\!\lesssim\!1$ & $d_5\!\leq\!\frac{5}{9}$ \\
\noalign{\vskip 0.5mm}
\hline
\noalign{\vskip 0.3mm}
\multicolumn{3}{l}{Predicted Modality$^\star$} & {\rm CU} & {\rm CU} & {\rm MLU} & {\rm CU} & {\rm CU} \\

\hline

\end{tabular}
\vskip0.10cm
~~~~~~~~~$^{\star}$CU: Certainly Unimodal; MLU: Most Likely Unimodal.


\end{table}

\subsection{Modality}\label{modal}

In a final set of diagnostic tests, we should also consider the modality of the mixture of the two derived Gaussian distributions (despite the fact that we can graph the combined distribution and see for ourselves whether two distinct modes emerge in the mixture). We consider 5 diagnostic test values of modality (for equal weights $p=1/2$) that we can obtain rather easily from the input parameters and the results listed in Table~\ref{t1} and in Table~\ref{t3} given below:

(1)~A likelihood ratio test for bimodality  \citep{rob69,hol08}, based on the value of $d_{1} = D_{\mu}\sqrt{{\cal V}/(D_1 D_2)}$, where $D_i=\sqrt{V_i\,}$ ($i=1, 2$). The mixture of the two Gaussians is unimodal if $d_{1}\leq 1$, or if
$$d_1>1~~~{\rm and}~~~\ln\left[d_{1}-(d_{1}^{~2}-1)^{1/2}{\mskip0.3\thinmuskip}\right] ~+~ d_{1}(d_{1}^{~2}-1)^{1/2} \leq 0\, .$$

(2)~Ashman's $d_2$ statistic \citep{ash94} in astrophysics, where $d_2 = 2 D_\mu (1 - D_\mu^{~2})^{-1/2}$. A clear separation of the components is expected when $d_2 \geq 2$, or equivalently, when $D_\mu > 1/\sqrt{2}$.

(3)~A bimodal separation index $d_3$ \citep{zha03}, defined by $d_3 = D_\mu[{\cal D}/(D_1+D_2)]$. This index moves above 1 only if the two Gaussian distributions essentially do not overlap.

(4)~The bimodality index $B_k=d_4$ of \cite{cha10}, based on variances and discussed in \S~\ref{significance} above. In our notation, it takes the form $d_4 = 1-D_\mu^{~2}$ (equation~(\ref{bk})), so if the separation distance between the two means is large, the index will assume low values.

(5)~A bimodality coefficient $d_5$ \citep{ell87}, based on the seminal work of \cite{pea16,pea29}, and defined by $d_5 = ({\cal S}^2 + 1)/{\cal K}$, where $d_5\in(0, 1)$ (cf. equation~(\ref{pearson_sk})). A bimodal distribution is expected to have light tails (low kyrtosis) and/or a strongly asymmetric form (high skewness), in which case the value of $d_5$ will be high. On the other hand, if $d_5\leq 5/9$ (the value for a uniform distribution), then the distribution is certainly unimodal \citep{Pfister2013}.

We calculated the above diagnostic quantities $d_j$ ($j=1$-5)  from the input parameters and the results listed in Table~\ref{t1} (and including additional values at each quarter phase from Table~\ref{t3} shown below). We collect these results in Table~\ref{t2}, where we also show the separation distances $D_\mu$ and $D_V$, the conditions for a unimodal distribution, and the predictions of the diagnostic quantities. All diagnostics indicate that the derived mixture is unimodal in each phase bin. 

Naturally, we have confirmed the above results by plotting the mixture of the two Gaussian distributions for each phase bin, which resembles the skewed \ion{He}{ii} 4686 lines observed in the spectra of IC10 X-1. The diagnostic analysis then serves to verify the robustness of the five criteria used for theoretical modality predictions.

\begin{figure}
\begin{center}
    \leavevmode
      \includegraphics[trim=0 0cm 0 0cm, clip, angle=0,width=9 cm]{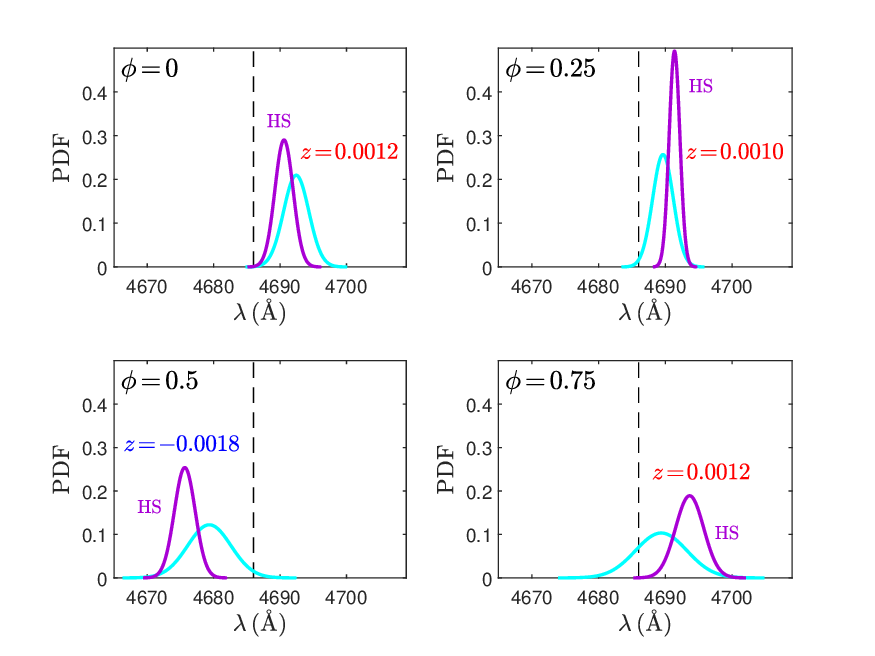}
\caption{Decomposed Gaussians for the skewed and kyrtic \ion{He}{ii} 4686 emission line of IC10 X-1. The two emitters switch places about their average wavelength every quarter phase. Their behavior is discussed in the text. Both component wavelengths shift to red or blue in unison. Average redshifts ($z>0$) and blueshifts ($z<0$) are noted in the frames, whereas individual component values are listed in Table~\ref{t5} below. In mid-X-ray eclipse, the hotspot is not directly observable; the average wavelength over the interval $\phi = 0.5\pm0.125$ is blueshifted by $-8.5$ \AA~relative to 4686 \AA; but the switching trend between the two components still occurs in this phase too.
\label{fig3}}
  \end{center}
\end{figure}

\subsection{Signal Decomposition to Two Gaussian Emitters}\label{decomp}

We can now follow the two distinct Gaussian components at various phases, as they move around and switch positions relative to one another. An illustration is shown in Figure~\ref{fig3}, where the two emitting components are shown in cyan and purple colors, respectively. The utilized spectra have been averaged in order to capture nearly equidistant phases separated by $\Delta\phi = 0.25$ (see also Table~\ref{t3} for the decomposed values corresponding to the four frames of the figure). Phases are roughly a quarter phase apart because stacking spectra does not produce exact quarter phases. This is noticeable in the quarter phases of Figure~\ref{fig3}, where the emission is systematically redshifted, with the implication that the average phase values are  $\phi\lesssim 0.25$ and $\phi\gtrsim 0.75$, respectively. The cyan component does however show blueshifted tails, especially at $\phi=0.75$. We have incorporated these approximations in Figure~\ref{fig:model} below, where we show schematic diagrams of the two emitting components around the binary orbit.

\begin{table}
\caption{\ion{He}{ii} Line Decomposition at Four Equidistant Phases}
\label{t3}

\addtolength{\tabcolsep}{-1pt}  
\hspace*{-\leftmargin}\begin{tabular}{rrrrrrrll}
\hline
\\
\multicolumn{5}{c}{Input Signal (Skewed \ion{He}{ii} Lines)} & & \\
\hline
$\phi$~~~ & $\mu^{\star}$ & ${\cal V}$~~~\, & ${\cal S}$~~~\, & ${\cal K}$~~~ &  &  \\
\hline
\noalign{\vskip 0.5mm}

0.00 &  5.5    &  3.601 &  0.349    &  3.064 \\
0.25 &  4.5    &  2.281 &  $-0.662$   &  3.231 \\
0.50 &  $-8.5$   &  9.971 &  0.721    &  3.274 \\
0.75 &  5.5    &  14.212&  $-0.623$   &  3.204 \\

\hline
\\

\multicolumn{7}{c}{Decomposition to Two Gaussians with Means $\mu_1^{} > \mu_2^{}$} \\
\noalign{\vskip 0.5mm}
\hline
$\phi$~~~ & $\mu_1^{~\star}$ & $\mu_2^{~\star}$ & $V_1$~~ & $V_2$~~ & $D_1$~~ & $D_2$~~ & ~~$_1$ & ~~$_2$ \\
\noalign{\vskip 0.5mm}
\hline
\noalign{\vskip 0.5mm}
0.00  & 6.43  & 4.59  & 3.62  & 1.89  &  1.90  & 1.38 & Wind & HS \\ 
0.25  & 5.37  & 3.65  & 0.65  & 2.41  &  0.81  & 1.55 & HS & Wind \\ 
0.50  & $-6.66$ & $-10.35$ & 10.66 & 2.47  &  3.27  & 1.57 & Wind & HS \\ 
0.75  & 7.64  & 3.38  & 4.45  & 14.92 &  2.11  & 3.86 & HS & Wind \\

\hline

\end{tabular}
\vskip0.15cm
~~~~~$^{\star}$(Line Center) $- ~(4686~\AA)$.

\addtolength{\tabcolsep}{1pt}  

\end{table}

\begin{table}
\label{t5}

\begin{tabular}{ccc}
\hline 
Phase & Shielded Wind Shift & Hotspot Shift \\
\,$\phi$ & $z_{_{\rm wind}}$ & $z_{_{\rm HS}}$ \\
\noalign{\vskip 0.5mm}
\hline
\noalign{\vskip 0.75mm}
0.00 & $1.37\times10^{-3}$  & $9.80\times10^{-4}$  \\
0.25 & $7.78\times10^{-4}$  & $1.15\times10^{-3}$  \\
0.50 & $-1.42\times10^{-3}$~~ & $-2.21\times10^{-3}$~~  \\
0.75 & $7.22\times10^{-4}$  & $1.63\times10^{-3}$   \\

\hline

\end{tabular}

\end{table}

The purple component is clearly more slender and taller than the cyan component, which is always spread out in $\lambda$-space (the area under each curve is equal to 1). These characteristics support the hypothesis that the two components originate in different regions of the WR wind outflow, and that they move independently  because they switch positions roughly every quarter phase. We believe that the extended (cyan) component originates in the expanding wind \citep[in the shadow of the WR star;][]{laycock2015revisiting}; whereas the slender (purple) component comes from a hotspot (HS) and the accretion stream impacting the accretion disk that has formed around the BH \citep{binder2021wolf}. This is because when the shielded wind expands along our line of sight (LOS) ($\phi=0$ and 0.5), we expect to see a large dispersion of velocities originating at various distances from the WR star. Specifically, we imagine the orbital configuration of the two emitters evolving as follows (see also Tables~\ref{t3} and~\ref{t5}, and the schematic diagrams drawn with precision in Figure~\ref{fig:model}):

\begin{itemize}
    \item[(a)]Phase $\phi=0.5$: Both components show maximum blueshifts and the wind velocities show a much larger dispersion ($D_{\rm wind}/D_{\rm HS}=2$; Table~\ref{t3}). The shielded wind is directed toward us, and the HS is located close to 9:00 on a 12-hour wall-clock centered on the BH (top view of Figure~\ref{fig:model}). Figure~\ref{fig:model} also shows that the BH is $10^\circ$ before coming to mid-eclipse because the average binary phase depicted in this panel is actually $\phi=0.44$.
    
    \item[(b)]Phase $\phi=0$: The shielded wind shows here maximum redshift (Table~\ref{t5}), as was expected. The slender stream/HS component shows a smaller redshift (Figure~\ref{fig3}). Then, the HS must be located close to 1:00 on its clock (Table~\ref{t5}: $\cos^{-1}(0.980/2.21)\simeq 64^\circ$, i.e., 2 hr behind the 3:00 mark, and no more than a few minutes off of the 1:00 mark). This configuration presents a $\sim$40\% larger dispersion of velocities, and a $\sim$40\% larger redshift to the shielded wind.

    \item[(c)]Phase $\phi=0.25$: The shielded sector of the wind is nearly orthogonal to the observer, yet it manages to produce a small mean redshift (Table~\ref{t5}). The HS shows about the same redshift as at $\phi=0$, thus it has not moved too far off from its previous location. Note however that its orbital period does not have to be 1:1 with the binary period; the HS could have done a full orbit returning back to about the same location. (See Appendix~\ref{appa} for an in-depth investigation of the orbital motion of the HS.) In this phase, the dispersion of wind velocities is about twice as much now as that of the HS; we note that the ratio D$_{\rm wind}$/D$_{\rm HS}$ is $\simeq$2 in the next two quarter phases as well (see Table~\ref{t3}, where the last two columns clarify the sources in each phase).
    
    \item[(d)]Phase $\phi=0.75$: The shielded sector of the wind is again nearly orthogonal to our LOS, and it manages again to produce a small mean redshift (Table~\ref{t5}). On the other hand, the HS shows its largest redshift ($\sim$74\% of the maximum blueshift), which places it close to 4:30 on its clock (Table~\ref{t5}: $\cos^{-1}(1.63/2.21) \approx 45^\circ$, i.e., 1.5 hr ahead of the 3:00 mark, and no more than 5 minutes off of 4:30). In this phase, characterized by the poorest quality spectra, we find only one striking difference relative to phase 0.25: the velocity dispersions of both the HS and the wind have more than doubled (D$_{0.75}$/D$_{0.25}\approx 2.5$, as obtained from the values listed in Table~\ref{t3}).\\
\end{itemize}

We summarize the counterclockwise movements of the HS around its clock during the BH orbital quarter phases from $\phi=0$ to $\phi=1$, respectively. In sequence, the HS is located roughly at 1:00, 0:50, 9:00, 4:30, and returns back to 1:00. Its orbit is somewhat eccentric as would be expected for an accretion stream that impacts the outer accretion disk of the BH. These timings allow us to determine approximately the axes of the ellipse, and dynamical theory allows us to work out the kinematics of the HS. We undertake this task in the next subsection, where our investigation comes to fruition.

\begin{figure*}
    \centering
    \includegraphics[width = 17.7 cm]{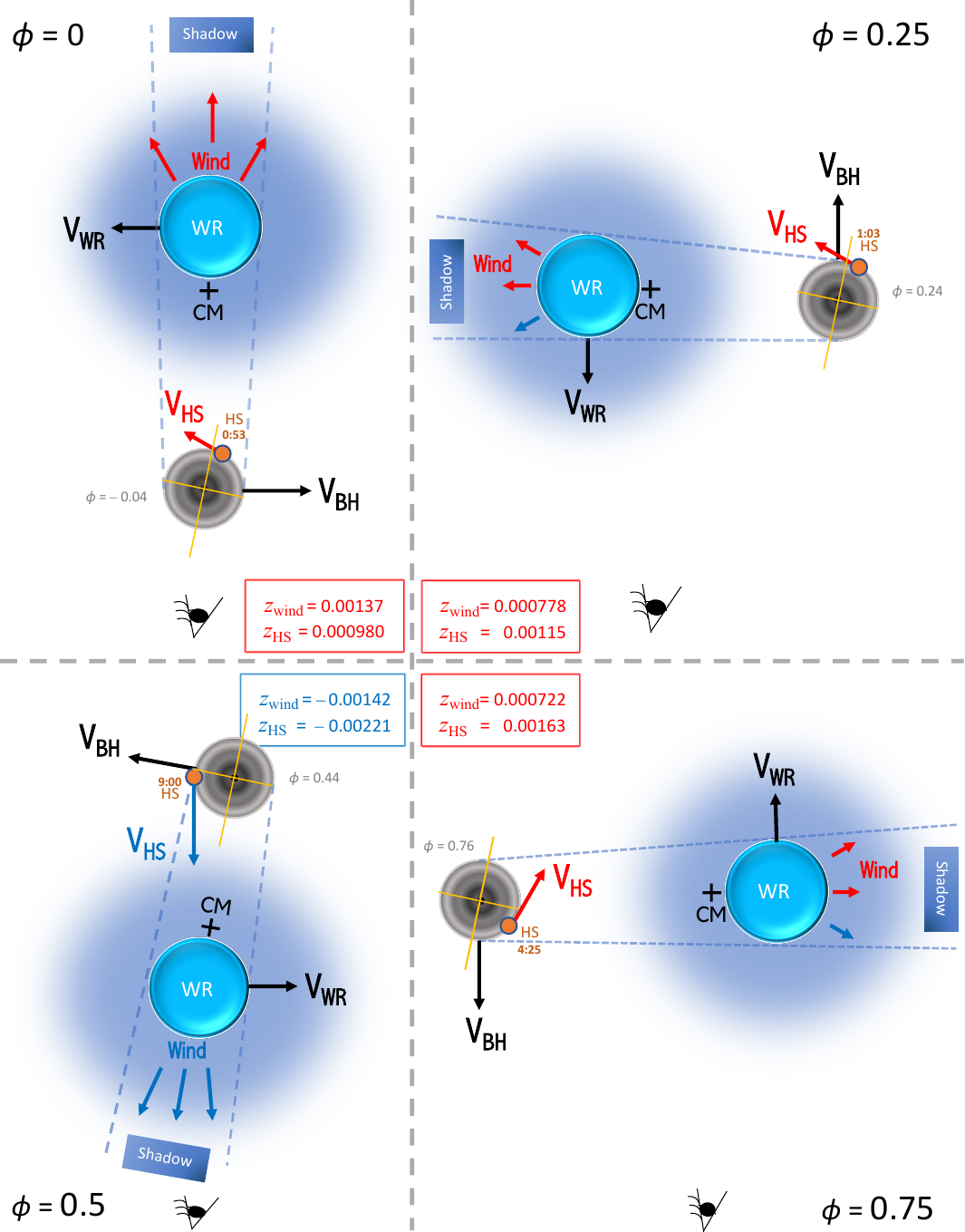}
    \caption{
    Schematic diagrams showing the motions of the two stars (WR star and BH), the shielded wind, and the hotspot (HS) at four orbital phases. All vectors and HS locations have been drawn to scale; the HS orbits counterclockwise (ccw) about the BH, as seen from above; the stars orbit ccw about their center of mass (CM); the velocities of the stars were scaled for a mass ratio of $M_{\rm BH}/M_{\rm WR}=0.5$. Average phase values for the BH around the CM are noted in gray color, and wall-clock times for the HS around the accretion disk are noted in brown color. The brown axes fixed to the disk denote the inclination of the elliptical stream/HS orbit to our LOS (the major axis is tilted by $-21^\circ$ from the 6:00-12:00 line).
    }
    \label{fig:model}
\end{figure*}

\begin{table}
\caption{Ellipse Geometry and Kinematics of the HS Orbit about the BH}
\label{tHS}

\begin{tabular}{rrrr}
\hline 
\noalign{\vskip 0.5mm}
Property & Symbol & Definition & Value \\
\noalign{\vskip 0.25mm}
\hline
\noalign{\vskip 0.75mm}
Rotation from LOS & $\psi$ & Negative Clockwise & $-21^\circ$ \\
Orbital Period & $P_{\rm HS}$ & $2\pi /\Omega^{\,\ast}$ & 6.96 \\
\\
Area & $A$ & $P_{\rm HS}/2$ & 3.48 \\
Areal Speed & $dA/dt$ & $h/2 = {\rm constant}^{\,\ast\ast}$ & 0.5 \\
Circumference & $C$ & $\pi(a+b)[1 + f(\xi)]^{\,\ast\ast\ast}$ & 6.614 \\
\\
Semimajor Axis & $a$ & $(r_{\rm max}+r_{\rm min})/2$ & 1.071 \\
Semiminor Axis & $b$ & $\sqrt{r_{\rm max}r_{\rm min}\,}$ & 1.035 \\
Ellipticity    & $\varepsilon$ & $b/a$ & 0.9665 \\
Eccentricity   & $e$ & $\sqrt{1-\varepsilon^2\,}$ & 0.2568 \\
Linear Eccentricity & $c$ & $ae$ & 0.275 \\
\\
\multicolumn{3}{r}{Distances $r$ from the Focus (BH)~~~~~~~~~} \\
$r-$periastron & $r_{\rm min}$ & $a(1-e)$ & 0.7957 \\
$r-$apastron & $r_{\rm max}$ & $a(1+e)$ & 1.3455 \\
$r-$covertex & $r_{\rm cov}$ & $a$ & 1.071 \\
\\
$r_{\rm max, min}-$ratio & $r_{\rm max}/r_{\rm min}$ & $(1+e)/(1-e)$ & 1.691 \\
$r_{\rm max, cov}-$ratio & $r_{\rm max}/r_{\rm cov}$ & $1+e$ & 1.257 \\
\\
Orbital Speeds & $V_r$ & $\sqrt{2/r - 1/a\,}$ & $V_r(r)$ \\
\noalign{\vskip 0.5mm}
$\dots$at periastron  & $V_{\rm max}$ & $V_r(r_{\rm min})=1+e$ & 1.257 \\
$\dots$at apastron    & $V_{\rm min}$ & $V_r(r_{\rm max})=1-e$ & 0.743 \\
$\dots$at covertices & $V_{\rm cov}$ & $V_r(a)=1/\!\sqrt{a\,}$ & 0.967 \\
\\
$V_{\rm max, min}-$ratio & $V_{\rm max}/V_{\rm min}$ & $r_{\rm max}/r_{\rm min}$ & 1.691 \\
$V_{\rm max, cov}-$ratio & $V_{\rm max}/V_{\rm cov}$ & $r_{\rm max}/r_{\rm cov}$ & 1.300 \\

\noalign{\vskip 0.75mm}
\hline
\end{tabular}

\vskip 0.75mm

$^{~\,\ast}\,\Omega$ is the angular velocity of the orbit (Appendix~\ref{appa}). \\
$^{\ast\ast}\,h\equiv 1$ is the specific angular momentum of the orbit. \\
$^{\ast\ast\ast}f(\xi) = 3\xi\left(10 +\!\sqrt{4-3\xi\,}\right)^{-1}$ \citep{ram14}, with $\xi\equiv(a-b)^2/(a+b)^2$ and an error of ${\cal O}(\xi^5)$. The same result, accurate to 3 decimal places can be obtained from $C\!=\!2A/r_{\rm H}$, where $r_{\rm H}$ is the harmonic mean of $a$ and $b$ \citep{chr16}.  \\

\centering
\begin{tabular}{rccccc}
\hline
\noalign{\vskip 0.5mm}
\multicolumn{6}{c}{Round-clock Time-stamps of the HS in each Binary Phase} \\
\noalign{\vskip 0.25mm}
\hline
\noalign{\vskip 0.75mm}
Binary Phases\, $\phi$: & 0.0 & 0.25 & 0.5 & 0.75 & 1.0  \\
\hline
\noalign{\vskip 0.5mm}
HS Motion: & \multicolumn{5}{c}{Counterclockwise} \\
\noalign{\vskip 0.5mm}
HS Time-stamps: & 1:03 & 0:53 & 9:00 & 4:25 & 1:03 \\
\noalign{\vskip 0.75mm}
Apsaidal Points: & & \multicolumn{1}{r}{~~~~0:42} & \multicolumn{2}{c}{6:42}\\
\noalign{\vskip -1.1mm}
 & & \multicolumn{1}{r}{~~~~Apastron} & \multicolumn{2}{c}{Periastron} \\
\noalign{\vskip 1mm}
Covertices: &  & \multicolumn{2}{r}{9:42~~~~~} & \multicolumn{2}{c}{3:42~~~~~~~~} \\
\noalign{\vskip -0.9mm}
 & & \multicolumn{2}{r}{Left~~~~~} & \multicolumn{2}{l}{\,~~Right} \\
\noalign{\vskip 0.25mm}
\hline

\end{tabular}

\end{table}

\subsection{The Eccentric Stream/Hotspot Orbit}\label{fruition}

The HS appears to be moving quite slowly (1:03$\to$0:53) within the first quarter of the orbital phase; and much faster within the third quarter (9:00$\to$4:25). These extremes allow us to determine the axes of the ellipse: the major axis is along the 0:42-6:42 line; periastron is at 6:42 for the orbiting HS, and apastron is at 0:42 due to the slow HS motion around the top of the clock; the minor axis then is along the 3:42-9:42 line that connects the covertices of the ellipse. Therefore, the apastron is tilted by $-21^\circ$ (i.e., clockwise) from 12:00. Since the HS clock does not rotate about the BH and the 6:00 mark always faces the observer, this angle is also the inclination of the apsidal line to our LOS.

The geometric properties of the above ellipse and the kinematics of the orbiting HS are collected in Table~\ref{tHS}. We use units such that the specific angular momentum $h$ of the HS and the standard gravitational parameter $GM_{\rm BH}$ \citep{bate71} are both equal to 1, and times are expressed in hours. This normalization scheme also causes the semilatus rectum $\ell = b^2/a$ \citep{cox69} to be 1, giving us the useful relation $a=b^2$ between the semiaxes $a$ and $b<a$ of the ellipse. The condition 
$$b<a\, ,$$ 
imposed by the kinematics (Doppler shifts) allows us to pinpoint the orbital period of the HS with surprising accuracy ($P_{\rm HS}=6.96$ hr; see Appendix~\ref{appa}). This value implies that the HS executes 5 full orbits about the BH for every full orbit of the BH (i.e., $1\frac{1}{4}$ orbits per quarter of the binary phase).

We also see in Table~\ref{tHS} that the eccentricity of the orbit is low ($b/a\approx 0.97$) and that the ratio of speeds 
$$V_{\rm max}\!:\!V_{\rm cov}\!:\!V_{\rm min}\,\,=\, 1.69\!:\!1.30\!:\!1\, .$$ 
Thus, the HS is orbiting faster(slower) by a factor of 1.3 as it traverses each of these locations on the upper(lower) half of the ellipse. As a consequence of Kepler's second law, this speed differential causes the HS to spend 2.91 hr between covertices when going around periastron ($0.418 P_{\rm HS}$), compared to 4.05 hr when going around apastron ($0.582 P_{\rm HS}$).

\section{Discussion}\label{discuss}

\subsection{Winds and Accretion Streams in HMXBs}\label{winds}

Relatively strong and broad emission lines, such as the \ion{He}{ii} 4686 line, appear in WR stars due to their pronounced stellar winds, and these spectral features most likely originate in regions within the outflowing winds, hence containing no tangible information about the binary orbits of such massive stars \citep{laycock2015revisiting,binder2021wolf}. So, it seems that the dynamical information is encoded in the powerful winds emanating from these stars (\S\S~\ref{decomp} and \ref{fruition}). The stellar wind structure is quite complex in WR stars because of its enormous spatial extent and the presence of a nearby compact accreting object (presumably a BH) that intensifies and complicates the dynamics and the emission processes in both optical and X-ray wavelengths.

\citet{castor1975radiation} and \citet{nugis2000mass} carried out theoretical investigations of stellar-wind velocities in isolated WR stars. Several other authors \citep{clark2004wolf,carpano200733,tutukov2016binary} studied the wind velocities of WR stars in HMXBs, such as IC10 X-1 and NGC300 X-1 \citep{binder2021wolf}. In the theoretical studies \citep[e.g.,][]{nugis2000mass}, the wind velocity $v_{\rm w}(r)$ at distance $r$ from the WR star is generally expressed by the relation 
\begin{equation}
    v_{\rm w}(r) = v_0 + (v_\infty - v_0)\left(1 - \frac{R}{r}\right)^{\,\beta} ,
\end{equation}
where $v_0$ is the outflow speed at the surface of the WR star, $R$ is its photospheric radius, and $\beta\approx 0.8$-1 \citep{castor1975radiation}. For the terminal (asymptotic) velocity $v_\infty$ in IC10 X-1, \citet{clark2004wolf} suggest a characteristic value of 1750 km~s$^{-1}$, corresponding to a rest-frame redshift of $z_{\rm CC}=5.83\times10^{-3}$. 

The largest rest-frame shift, $|z|_{\rm HS} = 2.21\times10^{-3}$, that we measured is the blueshift of the HS at 9:00 (Table~\ref{t5}; Figure~\ref{fig:model}), which is $\sim0.4 z_{\rm CC}$ corresponding to a typical speed of $\sim$700 km~s$^{-1}$. One may expect higher speeds in the accretion stream, in case the HS is not precisely located at 9:00 at orbital phase $\phi=0.5$ and a larger velocity vector is projected on to our LOS (Appendix~\ref{symmetries}); and also because of dispersion in the component velocities (Figure~\ref{fig3}). In any case, we do not expect to observe speeds much higher then $\sim$1000 km~s$^{-1}$ at optical wavelengths (Table~\ref{t7}). These assertions will be put to the test in  Appendix~\ref{unique} (see also Table~\ref{t6}), where we describe the unique determination of the orbital period of the HS about the BH.

\subsection{WR+BH Binary System IC10 X-1}

In the HMXB system IC10 X-1, the accretion process involves the capture of the stellar wind by the compact object and the creation of an accretion disk around the companion BH. The origin of the \ion{He}{ii} 4686 emission line detected in this system can be the shadowed sector of the stellar wind \citep{laycock2015revisiting}, or the accretion stream that impacts the outer accretion disk of the compact object \citep{binder2021wolf}, or both. In this study, we have decomposed the skewed and kyrtic \ion{He}{ii} emission from IC10 X-1 at various phases into two Gaussian components that presumably emanate from different wind regions around the binary \citep[and, apparently, not close to the stellar surface of the WR star, as was deduced by][]{bhattacharya2023probing}. 

Figure~\ref{fig3} and Table~\ref{t3} show that the two emitters switch positions relative to one another at different phases, as the skewness of the combined signal switches from positive to negative and back, roughly twice during each binary orbit (Figure~\ref{fig_sep}). The skewness of the \ion{He}{ii} 4686 line is created by the sleek (purple) component, as it dances around the more extended (cyan) component identified with wind emission. As the shadowed sector of the wind goes about its orbital phases, the purple component, identified with emission from the accretion stream/HS \citep{binder2021wolf}, executes its own orbital motion around the BH. 

A consistent model that shows these behaviors and precisely-scaled velocity vectors in each binary phase is drawn in Figure~\ref{fig:model}. In the four frames of the figure, we also assign to the HS accurate clock times (in brown color) drawn from a 12-hour wall-clock attached to the BH/accretion disk. These times help us describe the location of the HS in time, as it executes its own motion about the BH. So, for clarity, we use phase values for the binary orbit and BH-clock cycles for the HS. 

In the orbital quarter phases, where we expected nearly zero redshifts, the wind managed to show small nonzero redshifts, as compared to the "eclipse phase" in which both components are strongly blueshifted despite the average phase being $\phi=0.44$ (the BH is $\sim$10$^\circ$ off of superior conjunction with respect to the observer in the $\phi=0.5$ frame of Figure~\ref{fig:model}). But, to our surprise, the HS displays sizeable redshifts at the same times (in fact, maximum redshift at $\phi = 0.75$). It is obvious that this component is executing an independent motion about the BH and that its speed is varying along its orbit. In Table~\ref{t5}, we list the blueshifts (negative) and redshifts (positive) of the \ion{He}{ii} 4686 line that we have deduced around a full orbit of the binary. At $\phi=0$, we caught the shielded wind moving away from the observer (as was expected), at the same time that the HS shows a relatively small redshift. At $\phi\approx0.5$, with the BH at $10^\circ$ (i.e., 1 hr) before mid-eclipse, both components show maximum blueshifts.

\subsection{Hotspot Orbital Period and Distance from the BH}\label{HSorbit}

Imagine another 12-hour wall-clock overlaid on to the binary orbit and centered at the CM in Figure~\ref{fig:model}. The BH is at 6:00 (inferior conjunction) at $\phi=0$ and moves counterclockwise by 3 clock hours during each quarter of its phase. The overall orbit of the HS appears to be at least as fast as the binary orbit, implying a stream/HS period of $P_{\rm HS}=34.8$ hr \citep{carpano200733,laycock2015chandra} in the model depicted in Figure~\ref{fig:model}. But this does not have to be the actual period of the HS, and most likely it is not (see Appendix~\ref{appa}): During a phase change $\Delta\phi=0.25$, the HS can in principle complete any number of full cycles plus a fraction of a cycle that takes it to its next location. This extra fractional cycle is the only condition imposed on the HS by the results of the data analysis. In fact, the model cannot distinguish HS prograde versus retrograde rotation either: were the HS located at 3:00 on its own clock at $\phi=0.5$ and moving clockwise, the observed Doppler shifts would have been the same as those quoted in Figure~\ref{fig:model}.

Because we do not know with certainty the masses of the two stars in the binary \citep{laycock2015revisiting}; in order to make some progress; we had to carry out a separate analysis for the viable values of the orbital period of the accretion stream/HS and its distance from the BH. We summarize this dynamical analysis in Appendix~\ref{appa}, where we calculate these quantities assuming various stellar masses and relying on the insights gained from the model depicted in Figure~\ref{fig:model}. An important contribution to this investigation comes from an analysis of the time-dependent velocity dispersion of the stream/HS (phase-dependent standard deviations are listed in $\lambda$-space in Table~\ref{t3}), presumably caused by shear due to the tidal forces exerted by the BH itself.

\section{Summary of Results}\label{sum}

Our investigation of the \ion{He}{ii} 4686 emission line from the WR+BH binary system IC10 X-1 produced the following results:
\begin{itemize}
    \item[(1)]The \ion{He}{ii} 4686 line, although weak in the {\it Gemini/GMOS} spectra, is definitely skewed (skewness ${\cal S}\neq 0$) and kyrtic/curved (kyrtosis ${\cal K}>3$).
    
    \item[(2)]These asymmetric optical properties likely arise from a mixture of two approximate Gaussian emitters that do not emanate from the bulk of the WR star. We are convinced that one emitting component lies in the extended wind of the WR star, in particular, in the shadowed sector that is shielded from the X-rays emanating from the BH companion \citep{laycock2015revisiting}; whereas the other (always less dispersed) component originates from a HS in the BH's outer accretion disk, which is impacted by the backflowing accretion stream that develops from a stagnation point behind the BH  \citep{binder2021wolf}. 
    
    \item[(3)]Our decomposition of the overall signal into two Gaussian components reveals how these two emitting regions are evolving in time, as the binary is orbiting around its CM (Figures~\ref{fig3} and~\ref{fig:model}; Tables~\ref{t3} and~\ref{t5}). Figure~\ref{fig:model} depicts quite accurately that both components show maximum blueshift near mid-X-ray eclipse $\phi\approx0.5$---when the HS is at 9:00 on a 12-hour wall-clock fixed on to the BH/accretion disk for the sake of describing the motion of the stream/HS; and that only the wind shows maximum redshift at the BH's inferior conjunction ($\phi=0$) relative to the observer. 
    
    \item[(4)]The HS appears to be in a low-eccentricity orbit about the BH. At $\phi=0$, it is located near 1:00 on its clock (assuming counterclockwise motion), and, for this reason, it shows a relatively small redshift. On the other hand, it displays maximum redshift at $\phi=0.75$, after it has come around the clock to 4:30, on its way back to 1:00. Our efforts to investigate the kinematics and the dynamics of the stream/HS are described in \S~\ref{fruition} and Appendix~\ref{appa}. 
\end{itemize}

\section*{Acknowledgements}

We appreciate the comments and suggestions made by the referee that helped us produce a precise model of the binary system (Figure~\ref{fig:model} and Table~\ref{tHS}). We thank UMass Lowell and the Lowell Center for Space Sciences and Technology for supporting this research. This work was supported in part by NSF-AAG grant 2109004.

The research was also supported in part by the international \textit{GEMINI} Observatory, a program of NSF’s NOIRLab, which is managed by the Association of Universities for Research in Astronomy (AURA) under a cooperative agreement with the National Science Foundation, on behalf of the \textit{GEMINI} partnership of Argentina, Brazil, Canada, Chile, the Republic of Korea, and the United States of America. Our investigation was enabled by observations made from the {\it Gemini-North} telescope, located within the Maunakea Science Reserve and adjacent to the summit of Maunakea.

This investigation has made use of Astropy ({\tt www.astropy.org}) a community-developed core Python package and an ecosystem of tools and resources designed specifically for astronomy \citep{astropy:2013, astropy:2018, astropy:2022}.

\section*{Data Availability}

The raw data were downloaded from the {\it GEMINI} telescope data archive. The extracted spectra are available in this repository \citep{spec_data} in fits format. Additional processed data and products (extracted spectra in ASCII or pdf formats, tables, and figures) can be obtained by contacting the corresponding author.



\bibliographystyle{mnras}
\bibliography{skew_x1} 




\appendix
\section{Stream/HS Orbits and Dispersions for Various Assumed BH Masses and HS Periods}\label{appa}

We investigate dynamical scenarios for an outer accretion stream/HS orbiting the BH faster than the BH is orbiting the WR star \citep[which has a period of $P_{\rm orb}=34.8$ hr;][]{carpano200733,laycock2015chandra}; such as the unique solution with 
\begin{equation}
P_{\rm HS}=6.96~{\rm hr}\,,
\label{true_p}
\end{equation}
whose geometric properties and kinematics are listed in Table~\ref{tHS} above. This particular scenario and $P_{\rm HS}$-value are singled out from the two-component model depicted with the utmost precision in Figure~\ref{fig:model}, along with the dynamical considerations that follow.

\begin{table*}
\caption{IC10 X-1 HS physical characteristics in various $P_{\rm HS}$-scenarios (columns 2-4) with a BH mass of $17M_\odot$ \citep[such as the BH in IC10 X-1's "twin" X-ray source NGC300 X-1;][]{binder2021wolf} and $32M_\odot$ \citep{laycock2015revisiting}. In the last column, we display the ellipticity of the orbit of the HS about the BH.}
\label{t6}

\begin{tabular}{cccccccccc}
\hline 
  & & & & \multicolumn{3}{l}{~~\underline{HS Physical~ Characteristics}} & & & $\underline{\rm HS~Orbit~Geometry}$  \\
  & & \multicolumn{2}{c}{HS Cycles per BH~~~~~~} & & \multicolumn{2}{c}{\underline{$M_{\rm BH} = 17 M_\odot$}} & \multicolumn{2}{c}{\underline{$M_{\rm BH} = 32 M_\odot$}} & Ellipticity  \\
No. &$P_{\rm HS}$ & Quarter & Full & $\Omega$ & $R$  & $V$ & $R$ & $V$ & $b/a$  \\
 & (hr)\, & \multicolumn{2}{c}{Orbit} & (rad~s$^{-1}$) & ($R_\odot$) & (km~s$^{-1}$) & ($R_\odot$) & (km~s$^{-1}$)  \\
\noalign{\vskip 0.25mm}
\hline
\noalign{\vskip 0.75mm}
1 & 34.80~~&   ~~$\frac{1}{4}$ & 1 &    $5.0\times10^{-5}$&   13.86~~ &   ~~484  &
17.12~~ & ~~597 & 0.565  \\ 
\noalign{\vskip 0.6mm}
2 & 6.96&    $1\frac{1}{4}$ & 5 &    $2.5\times10^{-4}$&    4.74&      ~~827  &
5.85 & 1021 & 0.966  \\ 
\noalign{\vskip 0.6mm}
3 & 3.87&    $2\frac{1}{4}$& 9 &    $4.5\times10^{-4}$&    3.20&      1006  &
3.96 & 1242 & 1.176  \\ 
\noalign{\vskip 0.6mm}
4 & 2.68&    $3\frac{1}{4}$& 13 &    $6.5\times10^{-4}$&    2.51&      1137  &
3.10 & 1404 & 1.329  \\ 
\noalign{\vskip 0.6mm}
\hline

\end{tabular}

\end{table*}

\begin{table*}
\caption{IC10 X-1 HS physical and geometric characteristics of the best HS model 2, listed also in Table~\ref{t6}, and determined to be unique by the analysis conducted in Section~\ref{unique}. In the last two columns, we display the ellipticity and the eccentricity of the orbit of the HS about the BH.}
\label{t7}

\begin{tabular}{ccccccccccc}
\hline 
  & & & & \multicolumn{3}{l}{~~\underline{HS Physical~ Characteristics}} & & & \multicolumn{2}{l}{$\underline{\rm HS~Orbit~Geometry~(from~Table~\ref{tHS})}^{\,\ast}$}    \\
  & & \multicolumn{2}{c}{HS Cycles per BH~~~~~~} & & \multicolumn{2}{c}{\underline{$M_{\rm BH} = 17 M_\odot$}} & \multicolumn{2}{c}{\underline{$M_{\rm BH} = 32 M_\odot$}} & Ellipticity & Eccentricity  \\
No. &$P_{\rm HS}$ & Quarter & Full & $\Omega$ & $R$  & $V$ & $R$ & $V$ & $\varepsilon=b/a$ & $e=c/a$  \\
 & (hr)\, & \multicolumn{2}{c}{Orbit} & (rad~s$^{-1}$) & ($R_\odot$) & (km~s$^{-1}$) & ($R_\odot$) & (km~s$^{-1}$) &   \\
\noalign{\vskip 0.25mm}
\hline
\noalign{\vskip 0.75mm}
2 & 6.96&    $1\frac{1}{4}$ & 5 &    $2.5\times10^{-4}$&    4.74&      ~~827  &
5.85 & 1021$^{\,\ast\ast}$ & 0.9665 & 0.2568  \\ 
\noalign{\vskip 0.6mm}
\hline

\end{tabular}

\vskip 0.75mm

$^{\,\ast}$\, Major axis of the mild ellipse tilted clockwise from the LOS by $-21^\circ$. \\

$^{\,\ast\ast}$\,A $32 M_\odot$ BH mass does not appear to be viable, according to our observations; the circular HS speed it generates is too high (see also \S~\ref{winds}).  \\
\vskip 5mm

\centering
\begin{tabular}{cccccc}
\hline
\noalign{\vskip 0.5mm}
\multicolumn{6}{c}{Observed Velocity Dispersion $\Delta V$ and Associated Spread $\Delta R$ in the Stream/HS} \\
\noalign{\vskip 0.25mm}
\hline
\noalign{\vskip 0.75mm}
Binary & HS Clock & Trigonometric       &HS Speed~~~~\,& Vel. Dispersion & Rel. Half-Width$^{\,\star}$       \\
$\phi$ & Timestamp& Angle\,~($^\circ$)  &$V$~({\rm km~s}$^{-1}$)~~~~\,& $\Delta V$~({\rm km~s}$^{-1}$) & $(\Delta R)/R$   \\
\hline
\noalign{\vskip 0.5mm}
 0.0  & 0:53 & 64     &294~~~~\,& $\pm\,$44 &  $\pm\,$0.150   \\
 0.25 & 1:03 & 59     &344~~~~\,& $\pm\,$26 &  $\pm\,$0.075   \\
 0.5  & 9:00 & 180~~  &$662^{\,\star\star}$& $\pm\,$50 &  $\pm\,$0.076   \\
 0.75 & 4:25 & 318~\, &489~~~~\,& $\pm\,$67 &  $\pm\,$0.138   \\
\noalign{\vskip 0.25mm}
\hline

\end{tabular}

\vskip 0.75mm

$^{\,\star}$\,Half-width\, $|\Delta R| < 1 R_\odot$ using the $R$ values from model 2 at the top of the table. So, in both BH-mass cases covered in model 2, the stream extends out to $\leq$85\% of the BH Roche lobe. \\

$^{\,\star\star}$\,HS orbital speed $V=662$ km~s$^{-1}$ was observed near BH superior conjunction (also mid-X-ray eclipse). To make model 2 agree with this value, we could reduce the BH mass to $M_{\rm BH}=8.7 M_\odot$. Then, $M_{\rm BH}/M_{\rm WR}=0.256$, ~$a=18.8 R_\odot$, ~$(R_{L})_{\rm {\mskip0.5\thinmuskip}BH}=5.07{\mskip0.3\thinmuskip}R_\odot$, and $R = 3.79 R_\odot$, i.e., 75\% of the Roche lobe size. Alternatively, we could displace the HS from its 9:00 position at $\phi=0.5$ and determine its new location in each of the other three orbital phases. That action would also cause a rotation of the axes of the elliptical orbit (Section~\ref{symmetries}). \\

\end{table*}

\subsection{The Unique Orbit of the Stream/HS about the BH}\label{unique}

Table~\ref{t6} lists some possible periods and physical characteristics of the orbiting HS according to the diagram in Figure~\ref{fig:model} for a BH mass of $17M_\odot$ and $32M_\odot$, respectively. The possible number of HS cycles during each quarter binary phase ($\Delta\phi=0.25$) is shown in the second column of the table. Multiplication by 4 produces column 3, the number of complete HS cycles (1, 5, 9, 13) during one complete binary orbit. In the following calculations, we solve first for a roughly circular HS orbit around a BH using the famous Keplerian and kinematic equations
\begin{equation}
V = \sqrt[3]{G M \Omega\,}\, ,
\label{newt}
\end{equation}
and
\begin{equation}
R = \frac{V}{\Omega}\, ,
\label{vor}
\end{equation}
where $V$ and $R$ are the HS orbital speed and the mean distance from the BH, respectively, $G$ is the Newtonian gravitational constant, and 
\begin{equation}
\Omega\equiv\frac{2\pi}{P_{\rm HS}}\, ,
\end{equation}
is the angular velocity of the HS (column 3 in Table~\ref{t6}). 

Speed $V$ in Table~\ref{t6} characterizes the accretion stream/HS which is produced from a stagnation point behind the BH \citep{stevens1994stagnation} and the accelerated backflow of the gas returning toward the accretion disk \citep{huarte2013formation, el2019wind}. HS distance $R$ is given in units of the solar radius $R_\odot = 6.957\times10^5$~km; it should be compared to the separation of the two stars $a=20.0{\mskip0.5\thinmuskip}R_\odot$ and the volumetric size of the BH Roche lobe 
\begin{equation}
(R_{L})_{\rm {\mskip0.5\thinmuskip}BH}=6.41{\mskip0.3\thinmuskip}R_\odot\, ,
\label{RL}
\end{equation}
for a mass ratio of $M_{\rm BH}/M_{\rm WR}=0.5$ \citep[using the $r_{L}(q, a)$ formula of][]{egg83}; correspondingly, for the case with $M_{\rm BH}/M_{\rm WR}=1$ \citep{laycock2015revisiting} in Table~\ref{t6}, we find that $a=21.6{\mskip0.5\thinmuskip}R_\odot$ and 
\begin{equation}
(R_{L})_{\rm {\mskip0.5\thinmuskip}BH}=8.17{\mskip0.3\thinmuskip}R_\odot\, .
\label{RL2}
\end{equation}
Here, we also find that the WR star (for $R_{\rm WR}\approx 8 R_\odot$) fills its own Roche lobe ($R_{\rm WR}/(R_{L})_{\rm {\mskip0.5\thinmuskip}WR} = 0.98$), as compared to the former case in which $R_{\rm WR}/(R_{L})_{\rm {\mskip0.5\thinmuskip}WR} = 0.91$.

From the top row of Table~\ref{t6}, we see that an HS:BH periodicity of 1:1 (as in the illusion seen in Figure~\ref{fig:model}) is not acceptable since the HS distance $R\gtrsim 2(R_{L})_{\rm {\mskip0.5\thinmuskip}BH}$ in both cases, and the BH accretion disk/stream crosses well inside the Roche lobe of the WR star (for which $(R_{L})_{\rm {\mskip0.5\thinmuskip}WR}=8.80{\mskip0.3\thinmuskip}R_\odot$ and $(R_{L})_{\rm {\mskip0.5\thinmuskip}WR}=8.17{\mskip0.3\thinmuskip}R_\odot$, respectively). Thus, the HS has to be faster than the binary, and model 1 in Table~\ref{t6} is discarded.

Models 3 and 4 are also discarded, but for a different reason: In the last column of Table~\ref{t6}, we have also listed the ellipticity $b/a$ of orbit of the stream/HS about the BH (see \S~\ref{fruition}), as this was deduced from the nonequidistant clock timings of the HS in the binary orbital phases of Figure~\ref{fig:model}. Obviously, models 3, 4, and subsequent models with faster HS orbits must be discarded because they get the axis of the ellipse wrong. Therefore, we are left with only one viable model of the HS orbit, model 2 in Table~\ref{t6}.

\subsection{Properties of the Unique Model 2 Listed in Table~\ref{t6}}

The physical and geometric properties of model 2, the only viable model of the stream/HS orbiting about the BH, are summarized in Table~\ref{t7}. The most important dynamical characteristics of the stream and the BH have been pointed out by asterisks and have been discussed in the footnotes to the table. 

The geometric properties at the rightmost columns of the table indicate that the stream/HS orbit is only mildly elliptical. This allows us to not iterate on the circular model described in the main text (Figure~\ref{fig:model}) and make ellipticity corrections to the orbit of the HS. 

\subsection{Wrap-up of Procedures and the Two Unbroken Symmetries of the Model}\label{symmetries}

In the work that we described above, we undertook the following cumbersome steps concerning observations, data reduction, and data analysis and interpretation of the \ion{He}{ii} 4686 emission line emanating from the high-mass X-ray binary IC10 X-1:\\
\begin{itemize}
\item[(1)]We dealt with low-quality spectra of the source, the kind that most people would not consider useful at all.
\item[(2)]We devised a new technique to remove the continuum without losing the embedded weak signal.
\item[(3)]We devised a new analytical method that exploits skewness in the line profiles and separates the signal into two different Gaussian components. 
\item[(4)]We observed these components moving independently and dancing around one another (Figure~\ref{fig3}), as the BH+WR star binary kept orbiting, as usual, about its own center of mass. 
\item[(5)]We tried to interpret the motions of the two distributed components according to their mean Doppler shifts and their variances. We identified them with the shielded wind (large variances) blowing away from the WR star in the direction opposite to the BH, and with an accretion stream/HS (smaller variances) orbiting about the BH itself, just outside its accretion disk.
\item[(6)]We constructed a precise model of the motions of all emitting components during various orbital phases (Figure~\ref{fig:model}), and the model told us that the orbit of the HS is mildly elliptical (ellipticity $b/a\simeq0.97$).
\item[(7)]We determined the geometry and kinematics of the mildly elliptical orbit (Table~\ref{tHS}) and the orbital period of the HS about the BH ($P_{\rm HS}\simeq 7.0$ hr). 
\item[(8)]Finally, we analyzed various dynamical models of the stream/HS orbiting the BH, but we could not constrain the mass of the BH or the size of its accretion disk (Section~\ref{unique}).\\
\end{itemize}

We have learned to execute these steps with minimum hardship, and to obtain useful results that make physical sense. Our next target is IC10 X-1's "twin" X-ray source NGC300 X-1 for which high-quality spectra of the \ion{He}{ii} 1640 emission line are readily available \citep{binder2021wolf}, and the measured Dopper shifts are quite different than those in IC10 X-1.

Two obstacles still remain that prevent a categorical determination of the stream/HS orbit in the IC10 X-1 binary system:\\
\begin{itemize}
\item[(a)]We cannot resolve a prograde versus a retrograde motion of the stream/HS around the BH (\S~\ref{HSorbit}). In Figure~\ref{fig:model}, the HS is assumed to rotate in the counterclockwise direction about the BH, just as the BH does about the center of mass of the binary. This is prograde motion. For the retrograde motion of the HS, we relocate it from 9:00 to 3:00 on its clock at $\phi=0.5$ and imagine that it rotates clockwise. 
\item[(b)]We have no way of knowing whether the HS at $\phi=0.5$ is indeed close to 9:00, as depicted in Figure~\ref{fig:model}. We can place the HS anywhere on the left semicircle of its orbit, and it will still show a blueshift upon projection onto the LOS. Because the HS shows maximum Doppler shift at $\phi=0$, we assumed that its velocity vector is parallel to our LOS, thus we do not just see a large projection of an even larger invisible velocity vector.\\
\end{itemize}

The former symmetry creates ambiguity as to the location of the HS in each orbital phase.  But there exist only two possibilities. The latter symmetry creates many more ambiguities. For example, we can easily match the large velocities (827 and 1021 km~s$^{-1}$) of the two (very different) dynamical models listed in Table~\ref{t7} to the observed projected HS speed of 662 km~s$^{-1}$ at $\phi=0.5$ by relocating the HS 1hr+14min and 1hr+39min, respectively, away from 9:00 in either direction. The only good news in this conundrum is that the major axis of the elliptical orbit then rotates away from its $-21^\circ$ orientation to the LOS by only $\pm\!13^\circ$ and $\pm\!15^\circ$ for $M_{\rm BH}=17 M_\odot$ and $32 M_\odot$, respectively. 



\bsp	
\label{lastpage}
\end{document}